\documentclass[aps,prl,twocolumn]{revtex4}  % for review and submission
\usepackage{graphicx}  % needed for figures
\usepackage{dcolumn}   % needed for some tables
\usepackage{bm}        % for math
\usepackage{amssymb}
\newcommand{\degree}{\ensuremath{^\circ}}
\begin{document}

% The following information is for internal review, please remove them for submission
\widetext
%\leftline{Version 1.11 as of \today}
%\leftline{Primary authors: L. Casparis}
%\leftline{To be submitted to PRL}
%\centerline{INTERNAL DOCUMENT -- NOT FOR PUBLIC DISTRIBUTION}

% the following line is for submission, including submission to the arXiv!!
%\hspace{5.2in} \mbox{Fermilab-Pub-04/xxx-E}

\title{Evidence for Disorder Induced Delocalization in Graphite}
%\input author.tex   % input authors

%\centerline{author list dated 24 March 2010}
\author{Lucas Casparis\textsuperscript{1}}
\author{Andreas Fuhrer\textsuperscript{2}}
\author{Doroth\'{e}e Hug\textsuperscript{1}}
\author{Dominikus K\"{o}lbl\textsuperscript{1}}
\author{Dominik M.~Zumb\"{u}hl\textsuperscript{1}}
%\email{dominik.zumbuhl@unibas.ch}
\affiliation{\textsuperscript{1}Department of Physics, University of Basel, Klingelbergstrasse 82, CH-4056 Basel, Switzerland}
\affiliation{\textsuperscript{2}IBM Research, Z\"{u}rich Research Laboratory, S\"{a}umerstrasse 4, CH-8803 R\"{u}schlikon, Switzerland}
\date{\today}

\begin{abstract}
We present electrical transport measurements in natural graphite and highly ordered pyrolytic graphite (HOPG),
comparing macroscopic samples with exfoliated, nanofabricated specimens of nanometer thickness. The latter exhibit a very large c-axis resistivity $\rho_c$ -- much larger than expected from simple band theory -- and non-monotonic temperature dependence, similar to macroscopic HOPG, but in stark contrast to macroscopic natural graphite. A recent model of disorder-induced delocalization is consistent with our transport data. Furthermore, Micro-Raman spectroscopy reveals clearly reduced disorder in exfoliated samples and HOPG, as expected within the model -- therefore presenting further evidence for a novel paradigm of electronic transport in graphite.
\end{abstract}

\maketitle

%\section{\label{sec:level1}First-level heading}
% sections are not used for PRL papers

%This sample document demonstrates proper use of REV\TeX~4 (and
%\LaTeXe) in mansucripts prepared for submission to APS
%journals. Further information can be found in the REV\TeX~4
%documentation included in the distribution or available at
%\url{http://publish.aps.org/revtex4/}.

Graphite is a paradigmatic layered material and has been investigated intensively for many decades. The in-plane resistivity $\rho_{ab}$ is rather well described by a simple Drude model. However, the resistivity $\rho_{c}$ along the c-axis, perpendicular to the graphite basal planes, as well as its temperature dependence $\rho_c(T)$, are not described by the simple band structure model \cite{ono, sugihara}, and currently lack theoretical understanding despite extended efforts. The resistive anisotropy $R_A = \rho_c / \rho_{ab}$ is a convenient dimensionless parameter characterizing transport properties.

Carbon atoms in the graphite basal planes are strongly bound by covalent bonds, while much weaker Van der Waals forces bind the graphene sheets along the c-axis. Non-trivial disorder such as stacking faults and crystalline grains result in a mosaic angle and complicate electronic transport. For isotropic disorder, simple band theory \cite{ono} predicts $R_A = m_c/m_{ab}\sim\,$140, the ratio of the corresponding band masses. This agrees well with measurements in natural graphite (NG) \cite{primak, edman}. In highly oriented pyrolytic graphite (HOPG), the anisotropy was found to be much larger, even exceeding 10'000 in some experiments \cite{royal,tsuzuku}. Moreover, band theory \cite{ono} predicts a monotonic metallic temperature dependence for both $\rho_{ab}$ and $\rho_c$, resulting in a temperature independent anisotropy $R_A$. This is seen in NG \cite{spainbook}, but not in HOPG, where $\rho_c$ is non-monotonic with a maximum around 40\,K \cite{royal, tsuzuku, kempa}, similar to $\rho_c$ in other layered materials, such as NaCo$_2$O$_4$ \cite{valla} and Cuprates \cite{lavrov}. A large anisotropy far exceeding 100 combined with the non-metallic temperature dependence -- together referred to as anomalous behavior -- are currently not understood and present a fundamental problem in condensed matter physics \cite{spainbook,kelly,dresselhaus}.

\begin{figure}[h!]\vspace{-2mm}
    \includegraphics[scale=1]{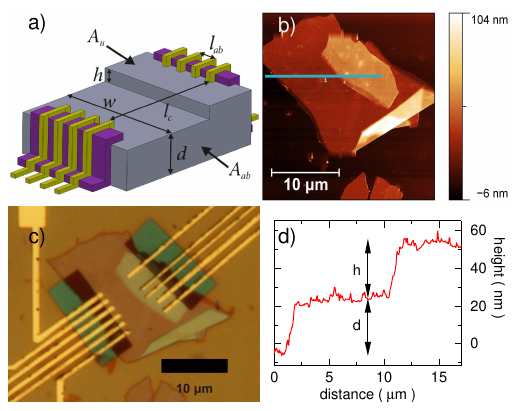}
    \caption{Nano-graphite samples. (a) Device schematic. Ti/Au contacts (yellow) for 4-wire measurements
    are patterned on each plateau, isolated from the graphite walls by a $SiO_2$ layer (purple). AFM picture
    (b) and optical microscope image (c) of an HOPG flake with two plateaus. (d) Cross section along the blue
    line in (b), giving plateau heights.}
    \label{device}
\end{figure}

In this work, we use exfoliation and nano-fabrication techniques to investigate both $\rho_{ab}$ and $\rho_c$ (see Fig.\,\ref{device}) in graphite flakes of various thickness in the nanometer range. Remarkably, we find in all types of graphite anomalous behavior -- namely a large resistive anisotropy as well as a non-metallic temperature dependence. Previous experiments measure $\rho_{ab}$ only \cite{esquinazi}. This permits a comparison of $R_A$ in samples with thicknesses in the nanometer range with macroscopic samples, which we also prepare and measure. The in-plane resistivity of all samples is in good agreement with reported values \cite{primak, edman, royal}, and shows no size-dependence. Therefore the large $R_A$ in the anomalous samples are to be attributed to a large $\rho_c$. The measured anisotropies appear consistent with a recent model
based on disorder induced delocalization by \emph{Maslov et al.} \cite{maslov, Zhang}, further corroborated by a disorder characterization of our samples using micro-Raman spectroscopy. Conduction path mixing due to a finite mosaic angle can account for the non-monotonic temperature dependence \cite{malsov2}, altogether presenting first experimental evidence for a novel paradigm of electrical transport in graphite.

To produce nanostep samples, we use the design shown in Fig.\,\ref{device}(a). We exfoliate graphite onto a
Si wafer with a $300\,$nm thick thermal oxide and identify suitable flakes with two plateaus differing in height by optical microscopy. The lower plateau height $d$ and the step height $h$ are determined from AFM images [see Fig. \ref{device}(b,d)], giving heights between 14 and $150\,$nm. To extend the range to larger step heights, we use e-beam lithography and oxygen-plasma etching to carve steps up to $h=450\,$nm.

For contacting the plateaus, we first cover parts of the exterior edges of both plateaus with $\mathrm{SiO_2}$ of at least $80\,$nm thickness \cite{NoteFab1} in order to prevent short-circuiting the c-axis. Contacts (typically a few hundred nanometers in width) and bonding pads are patterned in a final e-beam step, evaporating a Ti/Au layer thicker than $110\,$nm ($\mathrm{SiO_2}$ thickness plus $30\,$nm). A typical device is shown in Fig. \ref{device}(c). All resistances are measured with standard 4-wire lock-in techniques. This layout allows measurements of both $\rho_{ab}$ and $\rho_{c}$ on the same device, as needed to obtain the anisotropy. However, due to variations in the current distribution related to the individual device geometries, corrections to the measured resistances must be applied. We do this by means of a rough estimate based on the simplified geometry shown in Fig.\,\ref{device}(a). In addition, we also performed more elaborate numerical calculations of the current distribution to verify the observed effects \cite{SOM}.

\begin{figure}[h!]
    \centering
        \hspace*{-1.5mm}\includegraphics[scale=1.03]{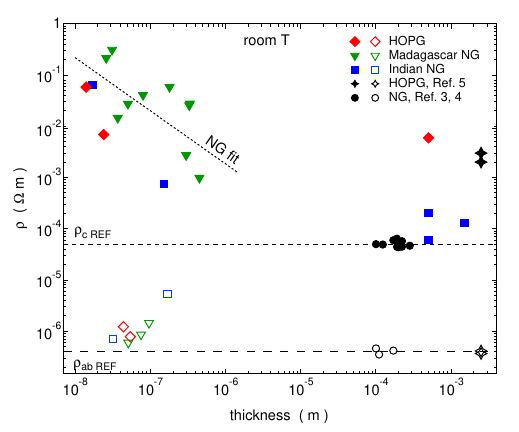}\vspace{-2mm}
    \caption{Influence of graphite thickness on $\rho_c$ (solid markers) and $\rho_{ab}$ (empty markers) at room
    temperature, comparing HOPG (red) with Madagascar NG (green) and Indian NG (blue). For $\rho_{ab}$, the abscissa value
    used is $d+h$, the overall flake thickness, see Table\,I in Ref.\,\cite{SOM}. Previous measurements of macroscopic samples (black) were added for both HOPG \cite{royal} (stars) and
    NG \cite{primak, edman} (circles) for comparison. Dashed horizontal lines indicate literature values
    $\rho_{ab\,\mathrm{REF}}$ for $\rho_{ab}$ and $\rho_{c\,\mathrm{REF}}$ for $\rho_c$. Further, the best power-law fit
    to all NG nanostep data yields an exponent of $-1.0\pm0.4$ and is shown by a dotted line to indicate a potential
    trend, see text.}
    \label{stepdep}
\end{figure}

The in-plane resistivity $\rho_{ab}=R_{ab} A_{ab}/l_{ab}$ is estimated from the 4-wire resistance $R_{ab}$ with current and voltage probes on the same plateau and assuming a simple rectangular shape of the graphite sample, with voltage probe distance $l_{ab}$ and total graphite cross section $A_{ab}$ [see Fig.\,\ref{device}(a)]. This is a good approximation for thin, elongated samples, small anisotropy and evenly distributed contacts. For realistic devices as the one shown in Fig.\,\ref{device}(b)+(c) and for large anisotropy, the extracted $\rho_{ab}$ presents an upper bound. Since the current cannot penetrate easily along the highly resistive c-axis and it's in-plane distribution is not homogeneous between the current contacts, the effective conduction channel is thinner and narrower than our estimate, thus reducing the actual resistivity below the estimate given here. Nevertheless, the $\rho_{ab}$ extracted (see Fig.\,\ref{stepdep} and Table\,I in Ref.\,\cite{SOM}) agree rather well with literature \cite{primak, edman, royal, tongay}. Moreover, $\rho_{ab}$ appears independent of the graphite thickness and is similar for NG (from two different sources, Indian NG and Madagascar NG) and HOPG samples, as seen in Fig.\,\ref{stepdep}, open symbols.

Next, we determine the c-axis resistivity $\rho_c$. Since $l_c$, the contact to contact distance across the step, is much larger than the step height $h$ (see Fig.\,\ref{device}(a)), we need to subtract the in-plane contributions to the measured resistance $R_{c}$ to obtain the actual c-axis resistance $\widehat{R_{c}}$ using
\begin{equation}
\widehat{R_{c}}=R_{c}-\rho_{ab}\cdot \left(\frac{l_{cl}}{w_l\cdot d}+\frac{l_{cu}}{w_u\cdot (d+h)}\right), \label{cax}
\end{equation}
with upper/lower contact to step distance $l_{cl/cu}$ and corresponding plateau widths $w_{l/u}$. $\rho_c$ is then
obtained from $\rho_c=\widehat{R_{c}}A_{u}/h$, where $A_{u}$ is the upper plateau area. Depending on the sample geometry, the in-plane correction can be a large fraction of $R_{c}$, see supplementary information for an overview. We note that as previously for $\rho_{ab}$, we again overestimate the thickness $d$ for large anisotropy. However, here, this tends to effectively cancel the overestimated $\rho_{ab}$, making the extracted $\rho_c$ quite robust. In order to test for the validity of the correction of Eq.\,\ref{cax}, we numerically calculate the current distribution for the various contact and sample geometries, taking into account the anisotropy. From a simultaneous fit of the two measurements of $R_{ab}$ and $R_{c}$ to the calculated resistances we can extract $\rho_{ab}$, $\rho_c$ and $R_A$ \cite{SOM}. As anticipated the simulated $\rho_{ab}$ are lower than the approximated $\rho_{ab}$. For both methods the $\rho_c$ values agree well with each other, corroborating our approach.

Figure\,\ref{stepdep} displays the resulting $\rho_{c}$ as a function of height (filled symbols), giving very large $\rho_c$ and correspondingly large anisotropy $R_A$ for all nano-graphites, both NG and HOPG. A power-law fit (linear fit on the log-log graph, slope $-1\pm0.4$) through all NG nanostep $\rho_c$ data points seems to indicate a trend of reduction of $\rho_c$ with increasing step height towards the macroscopic $\rho_c$ value in NG samples. HOPG nanostep data is excluded from the fit, since HOPG has no apparent size dependence when going from macroscopic to nanostep samples (filled red diamonds). In order to make a stronger statement, samples with step heights between $1\,\mathrm{\mu m}$ and $100\,\mathrm{\mu m}$ might give more insight \cite{NoteFab2}.

%\begin{table*}[htbp]
%\caption{\label{caxis} Nano-sample parameters, see text for definitions, and Fig.\,\ref{device} for an illustration.
%Samples listed here are represented in Figs.\,\ref{stepdep} and \ref{tdep}. Fig.\,\ref{stepdep} includes data from
%additional exfoliated Madagascar samples. }
%\begin{tabular}{cccccccrrcrcr} \hline
%Material &$\rho_{ab}$[$\mu\Omega m$]&$R_c[\Omega]$&$l_{cl}[\mu m]$&$w_{l}[\mu m]$&$l_{cu}[\mu m]$&$w_{u}[\mu m]$& $d [nm]$&$h[nm]$&$R_{corr}[\Omega]$&$A_u[\mu m^2]$&$\rho_c$[m$\Omega$$m$]&$R_{A}$\\
%\hline\hline
%HOPG 1       &  1.22 & 11.5 &4 &25&2.5&25& 30 &  14 &  2.3  & 361 & 58.0  & 47700 \\
%HOPG 2       &  0.78 & 12.1 & 6&20&2&20& 30 &  24 &  2.9  &  57 &  6.88 &  8840 \\
%\hline
%Madagascar 1 &  1.45 & 27.4 &1.2&10&2.3&10& 16 &  80 & 13.1  & 248 & 40.5  & 27900 \\
%Madagascar 2 &  0.60 & 26.0 &4.2&8& 3.3&5&14 &  37 &  5.6  &  95 & 14.5  & 24000 \\
%\hline
%India 1      & 0.71 & 18.9 &1&20&3&20& 15 &  17 & 10.7  & 100 & 63.2  & 89300 \\
%India 2      & 5.27 & 37.0 &2.4&20&2&15& 20 & 150 &  1.27 &  88 &  0.78 &   142 \\\hline
%\end{tabular}
%\end{table*}

\begin{figure}[htb]
    \centering
        \includegraphics[scale=1.03]{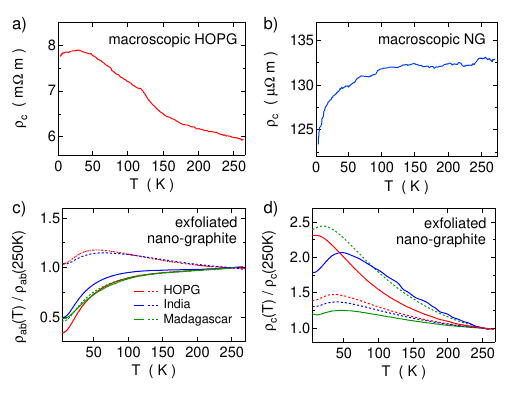}\vspace{-4mm}
    \caption{\vspace{-0mm}Temperature dependence of resistivities. $\rho_c(T)$ in macroscopic HOPG (a) and macroscopic Indian (b) NG.
    (c) $\rho_{ab}(T)$ in nanoscale samples for HOPG (red) and NG India (blue) and Madagascar (green). Two
    samples are presented for each graphite type (solid, sample 1; dashed, sample 2). For numerical values see Table\,I in Ref.\,\cite{SOM}.
    (d) $\rho_c(T)$ for the same samples.}\vspace{-4mm}
    \label{tdep}
\end{figure}

To allow a comparison with previous experiments, we also investigate macroscopic NG and HOPG samples, again measuring both $\rho_{ab}$ and $\rho_c$. Due to the geometry used, corrections due to a spreading of the current flow are small and not necessary for the macroscopic samples. On the other hand, the overestimation of the sample thickness due to a large anisotropy is still present, as in previous studies. The values obtained are also added to  Fig.\,\ref{stepdep}, together with typical values from literature \cite{edman, primak, royal}. We find decent agreement between our macroscopic data and previous measurements, reproducing here again the large discrepancy in $\rho_c$ between HOPG and NG in macroscopic samples.

Next, we turn to the temperature dependence $\rho_c(T)$ of the macroscopic samples \cite{ExpTemp}. For HOPG, we find a non-metallic $\rho_c$ at high $T$ ($d\rho_c/dT < 0$), see Fig.\,\ref{tdep} (a). Around $40\,$K, $\rho_c$ displays a rather shallow maximum, in good agreement with previous HOPG measurements \cite{royal}. In contrast, macroscopic Indian NG behaves weakly metallic and monotonic down to $4\,$K [see Fig.\,\ref{tdep} (b)], also in agreement with previous NG data \cite{primak}. Overall, our data from macroscopic samples fully agrees with the literature, indicating that our NG and HOPG is of comparable quality to that used in the literature, thus giving us confidence that a comparison of the exfoliated samples with literature is appropriate.

The temperature dependence of the exfoliated nano-graphite samples are shown in Fig.\,\ref{tdep} (c) and (d), normalized to the high-$T$ value. In most samples, $\rho_{ab}(T)$ is metallic and monotonous, as expected, and in agreement with macroscopic data \cite{primak, soule}. In two specimens, $\rho_{ab}$ exhibits a shallow maximum. This seems to occur occasionally in nanoscale samples, as previously reported \cite{kim, esquinazi}. Remarkably, $\rho_c(T)$ of all nanostep samples is qualitatively the same, showing a non-metallic and non-monotonic temperature dependence, qualitatively identical to macroscopic HOPG, and clearly different from the macroscopic NG data. We emphasize that the non-metallic $\rho_c(T)$ combined with the large anisotropy $R_A$ constitutes anomalous behavior for all nanoscale samples. In contrast, only macroscopic HOPG is anomalous, not macroscopic NG.

Motivated by an anisotropy far exceeding the band structure expectation, we consider a recent theory by Maslov et al. \cite{maslov}. A similar effect was also previously observed for photons \cite{Zhang}. Within this theory, c-axis transport is strongly suppressed in samples with weak bulk disorder due to 1D Anderson localization along the c-axis induced by randomly spaced barriers (e.g. stacking faults). This gives a very large $\rho_c$ and anisotropy $R_A$, in absence of strong bulk disorder. However, c-axis localization is destroyed by bulk scattering out of the c-axis direction, leading to reduced $\rho_c$ and smaller $R_A$. Interestingly, here, disorder can \emph{suppress} Anderson localization, rather than enhancing it, as is usually the case. Therefore, HOPG and nanostep samples are expected to have weak bulk disorder. In contrast, macroscopic NG specimens either have significantly more bulk disorder (suppressing c-axis localization), or fewer c-axis barriers, insufficient for localization (barriers spacing exceeding coherence length).

To characterize disorder, we measure spatially resolved micro-Raman spectra and use the D-peak intensity as a well established measure of graphitic disorder \cite{dresselhaus_2011,eren}. This peak arises from deviations from the ideal 2D graphene planes with sp$\rm{^2}$ bonds, reflecting various types of defects such as dislocations, impurities and other imperfections appearing in bulk graphite \cite{dresselhaus_2011}. Raman measurements present a characterization of the surface, since the laser ($\lambda$ = $532\,$nm) penetrates only about 50\,nm into the bulk graphite \cite{wang2008}. However, several of the nanosamples presented here have step heights less than or comparable to the laser penetration depth. Thus, for such samples, Raman spectroscopy reveals disorder throughout the entire volume of the relevant graphite sections where electron transport along the c-axis occurs -- offering a powerful tool to characterize disorder.

%whereas planar stacking faults and mosaic angles cannot be detected \cite{dresselhaus_2011}.

\begin{figure}[t]
    \centering
\vspace{-5mm}\hspace{-2mm}\includegraphics[scale=1.00]{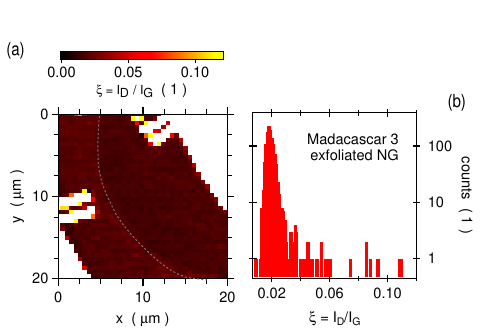}\vspace{-4mm}
    \caption{Disorder characterization with micro-Raman spectroscopy. (a) Ratio $\xi = I_D/I_G$ of the D-peak and the G-peak intensities scanned over an exfoliated Madacascar NG flake (Madagascar 3 \cite{SOM}, scan resolution $\sim 0.5\,\mathrm{\mu m}$), with 44\,nm tall  lower plateau (left) and 31\,nm step up (indicated by grey dashed curve) to the upper plateau (right) . White is off the graphite flake or on metal contacts. (b) Histogram of $\xi$ for the corresponding scan area.}
    \label{Raman}\vspace{-4mm}
\end{figure}

To quantify disorder, we introduce the intensity independent and dimensionless quantity $\xi=I_D/I_G$, the ratio of the D-peak intensity $I_{D}$ and G-peak intensity $I_G$ after subtraction of a constant background. For graphite, one finds $0\leq\xi\lesssim1$, where a large $\xi$ indicates a high degree of disorder. At clean locations, the D-peak disappears into the background and its measurement noise, giving $\xi\lesssim1/50$ corresponding to a D-peak indistinguishable from the noise. We have measured Raman maps of several nanostep samples. Figure\,\ref{Raman}(a) shows a typical scan, from one of the Madagascar samples, displaying only points showing the G-peak characteristic of graphite. We find that disorder is low $\xi\lesssim1/40$ essentially everywhere away from the edges or contacts, with D-peak within the measurement noise almost everywhere. This is seen on all exfoliated samples we have studied, and agrees well with the model expectations of low bulk disorder in nanostep samples.

The location of the step from the lower to the higher plateau -- indicated by the grey dashed curve -- leads to only a faint signature in $\xi$. This arises from a slight change in G-peak intensity at the step \cite{wang2008} and is not visible in the D-peak maps. This suggests that planar defects like stacking faults and grain boundaries may not be readily apparent in Raman scans. Within the model, such planar defects lead to c-axis localization and very large $\rho_c$ in nanosamples with weak bulk disorder as seen here. Thus, these observations are fully consistent with the model.

Exfoliating macroscopic NG even only once already results in strongly reduced surface disorder as seen here by Raman spectroscopy. This is already well known from scanning tunneling microscopy \cite{STM}. We have done a control measurements on macroscopic NG samples and find that $\rho_c$ remains low and essentially unchanged after one exfoliation step. Thus, surface disorder is not giving a significant contribution to $\rho_c$ in macroscopic samples.

Finally, we turn to the anomalous temperature dependence of $\rho_c$. If the c-axis resistivity $\rho_c$ is very large, the c-axis conductance path could easily be mixed with the ab-conductivity $\sigma_{ab}$ due to a finite mosaic angle $\theta$, effectively short circuiting the intrinsic c-axis conductivity $\sigma_c$. Assuming small tilting $\theta\ll1$, the measured conductivity $\widetilde{\sigma_c}$ can be written as \cite{malsov2}
\begin{equation}
\widetilde{\sigma_c}(T)=\sigma_{c}(T)+ \left\langle {\theta}^2 \right\rangle  \cdot \sigma_{ab}(T), \label{mechanism}
\end{equation}
\noindent where $\langle \theta^2 \rangle$ is the variance of $\theta$. In low bulk-disorder samples at low
temperatures, the intrinsic $\sigma_c$ is very small (strongly localized) and $\widetilde{\sigma_c}(T)$ obtains a significant component from $\sigma_{ab}$, including the (weakly) metallic temperature dependence $\sigma_{ab}(T)$, leading to a slight increase of $\rho_c(T)$ upon increasing $T$. At higher $T$, localization is weakened (due to phonon scattering, equivalent to increasing bulk disorder for increasing $T$), $\sigma_c$ is enhanced and becomes increasingly more dominant, leading to a decreasing $\rho_c$ above some cross-over $T$. For both HOPG and NG graphite we measure a mosaic angle between 0.2\degree and 2\degree (not shown), which is in agreement with the mixing mechanism, as a mosaic angle of about 0.8\degree corresponds to a cross-over $T$ of $\sim40\,$K. For disordered samples, on the other hand, the intrinsic $\sigma_c$ is dominating $\widetilde{\sigma_c}(T)$ since localization is already lifted by disorder, resulting in the usual metallic temperature dependence, as seen in macroscopic NG \cite{spainbook}.

This could potentially explain the size dependence mentioned in Fig.\,\ref{stepdep}: thinner samples tend to require more exfoliation steps, therefore becoming cleaner, more localized, and obtaining a larger $\rho_c$. Ultimately, for sufficiently small $h$, 1D localization should break down and $\rho_c$ decrease strongly -- not visible in the present data, presumably because $h$ is still too large. Overall, the Raman data is consistent with the predictions of the model, namely weak disorder in all exfoliated samples.

%neutron radiation experiments, more disorder, HOPG reduce \rho_c, NG increase \rho_c, as expected within this model.
%neutron irradiation inducing (bulk) disorder consistent with disorder induced localization model
%Raman penetration depth ca. 30 nm, measuring whole flake thickness for the thinner samples.

In conclusion, we investigate the anisotropy for nanoscale exfoliated graphite, and observe anomalous behavior, namely high $\rho_c$ and non-monotonic $\rho_c(T)$, in both NG and HOPG exfoliated samples. This is in stark contrast to macroscopic samples, where the anomalous behavior is only seen in HOPG, consistent with previous experiments. A recently proposed transport theory \cite{maslov} can consistently explain this convergence on the nanoscale, the macroscopic data, and the temperature dependence. This adds the nanoscale datapoints to the previously existing macroscopic graphite data. Furthermore, it is consistent with our finding of low disorder in exfoliated and HOPG samples, and high disorder in macroscopic NG. Notably, neutron irradiation experiments inducing bulk disorder have given consistent results \cite{royal, kelly}, i.e. reduced $\rho_c$ after irradiation of HOPG, further corroborating the model. We therefore present clear evidence of disorder induced delocalization, a conceptual novelty, as a new paradigm of electronic transport in graphite.

Though beyond the scope of the present work, it would be very interesting to subject the model to further scrutiny: studying intermediate steps filling the thickness gap in Fig.\,\ref{stepdep}, but also even smaller thicknesses, ultimately down to few- or bi-layer graphene, potentially revealing the localization length. This might be facilitated by bottom contacts with layers deposited on top, followed by top contacts. We note that the minimum thickness in the present samples is $14\,$nm, corresponding to about $50$ graphene layers. Further, a characterization of graphite disorder would be of great interest, e.g. investigating stacking faults and angles, intercalation, grain and boundary formation \cite{garcia}, aiming at identifying the localization mechanism, leading ultimately to a microscopic understanding of electrical transport in graphite. The results presented here were obtained in graphite, but it would be intriguing to see if similar arguments apply to other layered materials.

We are very grateful to D.~Maslov for initiating the experiments and numerous invaluable discussions and to S.~Tongay for performing XRD and Raman measurements. Further, we thank F.~Dettwiler, D.~Maradan and P.~Jurcevic for experimental help. This work was supported by the Swiss Nanoscience Institute (SNI), Swiss NSF, ERC starting grant and NCCR QSIT.


\begin{thebibliography}{99}
\bibitem{NoteFab1}%
  \BibitemOpen
  \bibinfo {note} {To accommodate the wide range of sample heights, we take the upper plateau height
plus $50\,$nm as the $\mathrm{SiO_2}$ layer thickness. For the thickest samples, angle evaporation was used
additionally to ensure complete coverage of the edges.}\BibitemShut{Stop}%
\bibitem{ExpTemp}%
  \BibitemOpen
  \bibinfo {note} {In a dip-stick, a calibrated Cernox chip-thermometer was mounted directly adjacent to the sample, ensuring a very
  similar temperature. The dip stick was moved slowly between liquid $^4$He and the room temperature top flange while continuously measuring
  both sample and thermometer, resulting in many points in $T$. Cool-down and warm-up curves were measured and compared, giving nearly
  identical results, and confirming the accuracy of this method. To obtain $\rho_c(T)$, we perform the correction following Eq. \ref{cax} at
every temperature, using $\rho_{ab}(T)$. }\BibitemShut{Stop}%
\end{thebibliography}
\end{document}